\documentclass[apj]{emulateapj}

\usepackage{subfigure,rotating,graphics,lscape,soul}

\def\deg{\hbox{$^\circ$}}              
\def\arcm{\hbox{$^\prime$}}            
\def\arcs{\hbox{$^{\prime\prime}$}}    
\def\lum{$\rm{erg}~\rm{s}^{-1}$}       
\def\lxs{$\rm{L}_{Xs}$}                 
\def\lxh{$\rm{L}_{Xh}$}                 
\def\lxt{$\rm{L}_{Xt}$}                 
\def\rsun{$\rm{R}_{\odot}$}           
\def\lsun{$\rm{L}_{\odot}$}           
\def\msun{$\rm{M}_{\odot}$}           
\def\lstar{$\rm{L}_{\star}$}          

\def\kms{km s$^{-1}$}                 

\def\emc{\hbox{EM$_1$}}              
\def\emh{\hbox{EM$_2$}}              
\def\lxb{$\rm{L_X}/\rm{L_{bol}}$}
\def\nhism{$\rm{N_{H}^{ISM}}$}
\def\nhl{$\rm{N_{H}^{local}}$}
\def\xmm{{\it XMM-Newton}}
\def\swift{{\it Swift}}
\def\na{New A}
\def\actaa{Acta Astron.}
\def\nar{New A Rev.}

\slugcomment{The Astrophysical Journal, 787:1, 2014}

\shorttitle{X-ray observations of WR 25}
\shortauthors{Pandey, Pandey \& Karmakar}

\begin{document}

\title{PHASE-RESOLVED \xmm ~AND \swift ~OBSERVATIONS OF WR 25}

\author{J. C. Pandey$^1$, S. B. Pandey, and  Subhajeet Karmakar}
\affil{Aryabhatta Research Institute of Observational Sciences (ARIES), 
Nainital-263 002, India; jeewan@aries.res.in}

\begin{abstract}
We present an analysis of long-term  X-ray and optical observations of the Wolf-Rayet binary WR 25.  Using archival data from observations with the XMM-Newton and the Swift observatories spanning over $\sim 10$ yr, we show that WR 25 is a periodic variable in X-rays with a period of  $208 \pm 3$ days.  X-ray light curves in the 0.5-10.0 keV energy band show phase-locked variability, where the flux increased by a factor of $\sim 2$ from minimum to maximum, being maximum near  periastron passage.   The light curve in the soft energy band (0.5-2.0 keV)  shows two minima indicating the presence of two eclipses.  However, the light curve in the hard energy band (2.0-10.0 keV) shows only one minimum during the apastron passage.  The X-ray spectra of WR 25 were explained by a two-temperature plasma model. Both the cool and the hot  plasmas were constant at $0.628\pm0.008$ and $2.75\pm0.06$ keV throughout an orbital cycle, where the cooler plasma could be due to the small scale shocks in a radiation-driven outflow and  the high temperature plasma could be due to the collision of winds. The column density varied with the orbital phase  and was found to be maximum  after the periastron passage, when the WN star is in front of the O star.  The abundances  of WR 25 were found to be non-solar. Optical V-band data of WR 25 also show the phase-locked variability, being at maximum near periastron passage.  The results based on the present analysis indicate that WR 25 is a colliding wind binary where  the presence of soft X-rays is attributed to individual components; however,  hard X-rays are due to the collision of winds.
  \end{abstract}


\section{Introduction}
Massive O-type stars evolve into Wolf-Rayet (WR) phases  when their hydrogen 
fuel has been consumed  and products of nuclear fusion appear in their 
atmosphere, before ending their lives as core-collapse supernovae 
\citep{doom87,smartt09,smartt09a}. Typically, the progenitors of WR stars  
have  initial  masses  greater than 25 \msun \citep{crowther07}, and they spend 
$\sim$10\% of their $\sim$ 5 Myr lifetime in the WR phase \citep{meynet05}. Spectra 
of WR stars are predominantly characterized by emission lines of He and N (WN 
stars), He and C (WC stars) and He and O (WO stars). In the evolutionary 
sequence of massive stars, WC and WO stars are expected to correspond to later 
stages than WN stars. WR stars are known to produce strong stellar winds driven 
by their strong radiation field. The stellar winds can reach velocities up to 
1000-3000 km s$^{-1}$ with mass-loss rates of $10^{-4 ~\rm{to} ~-6}$ \msun 
yr$^{-1}$, depending upon  mass and age  \citep{hamann06}. These 
winds not only affect the evolution of WR stars but also have a tremendous 
impact on their ambient media.  The detailed  physical properties of WR stars 
are summarized in many past reviews  \citep{abbott87,vanderhucht92,crowther07}.

X-ray emissions from WR binaries may consist of the combined emission from 
intrinsic stellar wind shocks and colliding wind shocks  between the two binary 
components \citep{prilutskii76,cherepashchuk76,luo90,usov92,stevens92}. 
Colliding wind binary systems often exhibit periodic X-ray modulation either 
because of the  change in binary separation  or due to the changing 
circumstellar opacity along the line of sight to the collision zone, resulting 
from the orientation of the system with respect to the observer.  Serious 
investigations of these variations were carried out after the advent of  
high-quality facilities like \xmm~ and {\it Chandra}.
Many O+O and WR+O  binaries as  observed with \xmm , show different kinds  of 
phase-locked modulations e.g., V444 Cyg and CD Cru \citep{bhatt10a},  HD 159176 
\citep{debecker04}, HD 152248 \citep{sana04}, HD 93403 \citep{rauw02}, etc.  It 
is believed that single WR and OB stars  emit soft X-rays (kT $<$ 1 keV) via 
shocks that are set up by instabilities in their supersonic line-driven winds 
\citep{lucy80,lucy82,owocki88}.  \cite{grafner05} have shown that  winds of 
WR stars can be driven by radiation pressure alone if multi-scattering effects 
are taken into account.

WR 25 (=HD 93162)  is a bright (V $\sim$ 8.03 mag) WR binary  located in the 
Carina Nebula region. WR 25 is classified as WN6h + O4f \citep{vanderhucht01}. 
\citet{gamen06} detected periodic radial velocity variations in WR 25 and  
suggested  that  WR 25 is an eccentric binary system with a probable period of 
about 208 days. The basic parameters of WR 25 are summarized in Table 1. The origin 
of the large X-ray flux of  WR 25  is suggestive of colliding wind emission in a 
binary system \citep{pollock87,raassen03,pollock06}. The  X-ray to bolometric 
luminosity (\lxb) ratio of $\sim 10^{-5.7}$ \citep{seward82} for WR 25 is an order of 
magnitude higher than those observed  for  single massive stars.
\cite{raassen03}  noticed that the X-ray luminosity of WR 25 had remained 
relatively  constant over a time span of 10 yr. Later, \citet{pollock06} found 
an increase in X-ray luminosity of more than a factor of two and suggested that 
the observed X-ray variability is  a result of colliding wind emission in a 
moderate eccentric binary.   In this context, we use the full set of 
archival XMM-Newton and Swift X-ray observations of WR25 to further investigate 
the properties of the colliding winds in this system. Our aim also extends to 
search for variability in the V-band using the All Sky Automated Survey
 [ASAS; \cite{pojmanski02}] archival data. 

The paper is organized along the following lines: Section \ref{sec:observation}, 
describes the observations and data reduction. The light curve analysis is given 
in Section \ref{sec:lightcurve}, X-ray spectral properties of WR 25 are 
described in Section \ref{sec:spectra}. Section \ref{sec:optical} describes the 
V-band optical observations. In Section \ref{sec:discussion}, we present  
the discussion and conclusions.

\begin{table}
\caption{Basic parameters of WR+O binary, WR 25.}
\label{tab:para}
\tabcolsep=0.14cm
\begin{tabular}{p{1.7cm}p{1.6cm}p{0.3cm}p{1.9cm}p{1.0cm}p{0.3cm}}
\hline
\hline
Parameters  & Value & Ref$^a$ &Parameters  & Value & Ref$^a$\\
\hline
Period (d)       &$ 207.85\pm0.02$&1&$v_\infty$(\kms) & 2480&2\\
$e$ (\deg)       &$ 0.50\pm0.02$  &1&$\dot{M}$(M$_\odot$/yr) &10$^{-4.3}$&2 \\
V$_0$ (\kms)     &$ -34.6\pm0.5$  &1& dist.(kpc) & 3.24  & 3\\
K (\kms)         &$ 44\pm2$       &1& \lstar(\lsun)& 10$^{6.2}$ &2 \\
$\omega$ (\deg)  &$ 215\pm3$      &1& E(B-V) (mag)& 0.63 & 3 \\
T$_0$ (HJD)      &$ 2451598\pm1$  &1& \lxb& $10^{-4}$&4 \\
$a~sin~i$ (\rsun)&$ 156\pm8$      &1& ...&...&...\\
\hline
\end{tabular}
~~~\\
Here: $e$ - eccentricity, V$_0$ - radial velocity, K - radial velocity 
amplitude, $\omega$ - orientation of periastron, T$_0$ - Julian date of 
periastron passage, $ a~sin~i$ - semi major axis, $v_\infty$ - terminal 
velocity, $\dot{M}$ - mass loss rate\\
$^a$References: (1) \cite{gamen06}; (2) \cite{crowther98};  (3) 
\cite{vanderhucht01}; (4) \cite{seward82} \\

\end{table}

\section{Observations and data reduction}
\label{sec:observation}
\subsection{XMM-Newton}
 WR 25 was observed with the \xmm ~ satellite using various detector configurations
on twenty occasions  from the year 2000 to 2009, spanning  $\sim$8.5 yr.
The \xmm ~ satellite is composed of three coaligned X-ray telescopes
\citep{jansen01}, which simultaneously observe a source, accumulating
photons in three CCD-based instruments, namely  the nearly-identical MOS1 and MOS2 \citep{turner01}
detectors  and the PN \citep{struder01} detectors, which comprise the European Photon 
Imaging Camera (EPIC). The EPIC instrument provides  imaging and spectroscopy 
in the energy range from 0.15 to 15 keV with an  angular  resolution of 4.5-6.6 
arcsec  and a spectral resolution ($E/\Delta E$) of 20-50. Exposure times for
 these observations  were in the range of 6-60 ks.  A log of observations is 
provided in Table \ref{tab:xmmlog}.

The data were reduced with standard \xmm ~Science Analysis System ({\sc SAS})
software, version 12.0 using version 3.1 calibration files. The 
pipeline processing of raw EPIC Observation Data Files was done using the
{\sc epchain} and {\sc emchain} tasks which allow calibrations both in energy 
and astrometry of the events registered in each CCD chip.
We have restricted our analysis to 
the energy band 0.5 - 10.0 keV as the background contribution is particularly 
relevant at high energies where stellar sources have very little flux and are 
often undetectable. Event list files were extracted using the {\sc SAS} task 
{\sc evselect}. The {\sc epatplot} task was used for checking for 
pile-up effects and  no observations were  affected by  pile-up.
 Data from the three cameras were individually screened for the time 
intervals with high background. The observations affected by high background 
flaring events were excluded (see Table 2). X-ray light curves and spectra of 
WR 25 were generated from on-source counts obtained from circular regions with a
radius of 30\arcs ~around the source. 
The background was chosen from several source-free regions on the detectors   
surrounding the source.  We used the tool {\sc epiclccorr} to correct for good 
time intervals, dead time, exposure, PSF, quantum efficiency and  background
subtraction. The SAS task {\sc especget} was used to generate the spectra, which 
also computes the photon redistribution as well as the ancillary matrix. Finally, 
the spectra were rebinned to have at least 20 counts per spectral bin.

\subsection{SWIFT}
WR 25 has also been regularly monitored by the \swift ~X-ray telescope (XRT) 
since 2007 November.  The XRT observes from 0.3 to 10 keV using CCD detectors, 
with energy resolution of $\approx 140$ eV  at $\sim 6$ keV \citep{burrows05}. 
The exposure times for the XRT observations varied from 1 to 26 ks. All XRT data 
were collected in photon counting mode. A log of observations is given in Table 
\ref{tab:swiftlog}.  We have excluded observations with exposure times less than 
2 ks due to poor or no signal.  In order to  produce the cleaned and calibrated 
event files, all the data were reduced using the \swift ~ {\sc xrtpipeline} task 
(version 0.12.6)  in which standard event grades of 0-12 were selected, and 
calibration files  were used from the CALDB 2.8 
release\footnote{http://heasarc.gsfc.nasa.gov/docs/heasarc/caldb/caldb\_intro.html}.

 For every observation, images, light curves and spectra were obtained with the 
{\sc xselect} (verson 2.4) package. For each observation, source spectra and light 
curves  were extracted from a circular region with a radius of 30\arcs. For the 
background estimation, an annular region with inner and outer radii of 69\arcs 
~and 127\arcs  ~were used around the source region. The spectra were grouped to 
have a minimum of twenty counts per energy bin with {\sc grppha}. The spectra 
were corrected for the fractional exposure loss due to bad columns on the CCD. 
For this, we created exposure maps with the {\sc xrtexpomap} task, which is used 
as an input to generate the ARF with the {\sc xrtmkarf} task. For the RMF, the 
latest version was used from  the HEASARC calibration data base.

\begin{table*}
\caption{Log of observations of WR 25 from \xmm.} \label{tab:xmmlog}
\begin{tabular}{ccccccrcrc}
\hline
\hline
Data&Rev.&Observation&Detector$^1$ & Obs. Date  &U.T.$^3$  & Exposure& Bkg flare$^4$ & Effective  &Off axis \\
set &~   & ID        & (Filter$^2$)&(YYYY-MM-DD)&(hh:mm:ss)    & Time(s)& from-to (ks) & exposure(s)&(\arcm)\\
\hline
x01 &  115  & 0112580601$^5$&M1(TH)  &2000-07-26   &  05:07:47  &   33589  & 
32.8-33.6       &32761&     6.864\\
    &       &               &M2(TH)  &             &  06:00:45  &   30488  & 
29.6-30.5       &29585&\\
    &       &               &PN(TH)  &             &  05:48:51  &   31198  & 
30.3-31.2       &30298&    \\
x02 &  116  & 0112580701$^5$&M1(TH)  &2000-07-27   &  23:57:54  &   10991  &      
-          &10991&     6.877\\
    &       &               &M2(TH)  &2000-07-28   &  00:50:51  &    7991  &      
-          & 7991&               \\
    &       &               &PN(TH)  &2000-07-28   &  00:38:57  &    8999  &      
-          & 8999&\\
x03 &  283  & 0112560101$^5$&M1(TH)  &2001-06-25   &  07:34:37  &   34249  &   
0-9.6         &24681&     0.066	\\
    &       &               &M2(TH)  &             &  07:40:30  &   33894  &   
0-9.6         &24325&           \\
    &       &               &PN(TH)  &             &  06:51:26  &   33133  &   
0-8.7         &24433&             \\
x04 &  284  & 0112560201$^5$&M1(TH)  &2001-06-28   &  07:29:20  &   35886  & 
30.0-35.9       &30000&     0.094\\
    &       &               &M2(TH)  &             &  07:29:20  &   35890  & 
30.0-35.9       &30000&            \\
    &       &               &PN(TH)  &             &  07:22:56  &   37364  & 
27.7-37.4       &27733&          \\
x05 &  285  & 0112560301$^5$&M1(TH)  &2001-06-30   &  04:45:30  &   37107  & 
33.6-37.1       &33636&     0.066\\
    &       &               &M2(TH)  &             &  04:45:28  &   37108  & 
33.6-37.1       &33636&              \\
    &       &               &PN(TH)  &             &  05:23:05  &   34526  & 
31.4-34.5       &31381&                \\
x06 &  573  & 0145740101$^6$&M1(TH)  &2003-01-25   &  12:58:20  &    6812  &     
-           & 6812&     7.912\\
    &       &               &M2(TH)  &             &  12:58:20  &    6813  &     
-           & 6813&                       \\
x07 &  574  & 0145740201$^6$&M1(TH)  &2003-01-27   &  01:03:43  &    6961  &     
-           & 6961&     7.855\\
    &       &               &M2(TH)  &             &  01:03:43  &    6961  &     
-           & 6961&   \\
x08 &  574  & 0145740301$^6$&M1(TH)  &2003-01-27   &  20:37:04  &    6812  &     
-           & 6812&     7.856\\
    &       &               &M2(TH)  &             &  20:37:05  &    6812  &     
-           & 6812&   \\
x09 &  575  & 0145740401$^6$&M1(TH)  &2003-01-29   &  01:40:32  &    8311  &     
-           & 8311&     7.824 \\
    &       &               &M2(TH)  &             &  01:40:33  &    8311  &     
-           & 8311&    \\
x10 &  575  & 0145740501$^6$&M1(TH)  &2003-01-29   &  23:55:06  &    6811  &     
-           & 6811&     7.815\\
    &       &               &M2(TH)  &             &  23:55:07  &    6811  &     
-           & 6811&    \\
x11 &  640  & 0160160101$^7$&M1(TH)  &2003-06-08   &  13:30:32  &   28415  
&0-1.4,8.9-       &12735&     5.965  \\
    &       &               &    &             &            &          & 
-10.3,15.5-28.4 &     &            \\
    &       &               &M2(TH)  &             &  13:30:22  &   28414  &  as 
above       &12735& \\
x12 &  662  & 0145780101$^6$&M1(TH)  &2003-07-22   &  01:51:36  &    8300  &       
-         & 8300&     6.335\\
    &       &               &M2(ME)  &             &  01:51:29  &    8300  &       
-         & 8300&    \\
x13 &  668  & 0160560101$^6$&M1(TH)  &2003-08-02   &  21:01:14  &   17649  & 
4.7-13.7        & 8649&     6.516\\
    &       &               &M2(ME)  &             &  21:01:03  &   17667  & 
4.7-13.7        & 8649&                \\
x14 &  671  & 0160560201$^6$&M1(TH)  &2003-08-09   &  01:44:20  &   12652  & 
9.5-10.3        &11863&     6.557\\
    &       &               &M2(ME)  &             &  01:44:08  &   12652  & 
9.5-10.3        &11863&              \\
x15 &  676  & 0160560301$^6$&M1(TH)  &2003-08-18   &  15:23:34  &   18850  &      
-          &18850&     7.070\\
    &       &               &M2(ME)  &             &  15:23:22  &   18850  &      
-          &18850&       \\
x16 & 1126  & 0311990101$^7$&M1(TH)  &2006-01-31   &  18:04:22  &   63034  &      
-          &63034&     7.656   \\
    &       &               &M2(TH)  &             &  18:04:20  &   66036  &      
-          &66036&     \\
    &       &               &PN(TH)  &             &  18:26:30  &   30000  &    
0-3.0        &26961& \\
x17 & 1662  & 0560580101$^8$&M1(TH)  &2009-01-05   &  10:23:02  &   14602  &      
-          &14602&     8.025\\
    &       &               &M2(TH)  &             &  10:22:53  &   14602  &      
-          &14602&      \\
x18 & 1664  & 0560580201$^8$&M1(TH)  &2009-01-09   &  14:28:07  &   11603  &      
-          &11603&     7.999\\
    &       &               &M2(TH)  &             &  14:27:58  &   11603  &      
-          &11603&     \\
x19 & 1667  & 0560580301$^8$&M1(TH)  &2009-01-15   &  11:22:56  &   26552  &      
-          &26552&     7.942\\
    &       &               &M2(TH)  &             &  11:22:46  &   26552  &      
-          &26552&\\
x20 & 1676  & 0560580401$^8$&M1(TH)  &2009-02-02   &  05:31:18  &   23903  &      
-          &23903&     7.757\\
    &       &               &M2(TH)  &             &  04:46:09  &   26613  &      
-          &26613&\\
    &       &               &PN(TH)  &             &  04:45:24  &   24669  &      
-          &24669&       \\
\hline
\end{tabular}
~~\\
{$^1$} M1 and M2 stand for MOS1 and MOS2, respectively.
$^2$ TH and ME stand for thick and medium filters, respectively.
{$^3$} Exposure start time.  {$^4$} Background proton flare duration during the 
observations.\\
PIs of observations were {$^5$}Dr. Albert Brinkman, {$^6$}Dr. Michael Corcoran, 
{$^7$}Dr. Fred Jansen, and {$^8$}Dr. Kenji Hamaguchi
\end{table*}


\begin{table}
\caption{Log of observations of WR 25 from \swift.} 
\label{tab:swiftlog}
\begin{tabular}{p{0.3cm}p{1.7cm}p{1.9cm}p{1.4cm}p{0.8cm}p{0.6cm}}
\hline
\hline
Set&Observation&Obs. Date&U. T.&Exp.&offset\\
&ID&(YY-MM-DD)&(hh:mm:ss)&Time(s)&(\arcm)\\
\hline
s01 & 00037049001$^1$ &  2007-12-11& 14:09:00  &  4490.5&  6.339 \\
s02 & 00037049002$^1$ &  2008-01-19& 06:08:01  &  3246.6&  4.682 \\
s03 & 00031097001$^2$ &  2008-01-25& 02:00:01  &  4440.5&  1.342 \\
s04 & 00031097002$^2$ &  2008-01-26& 02:10:01  &  3717.2&  0.387 \\
s05 & 00031097003$^2$ &  2008-01-27& 00:44:00  &  3452.8&  3.892 \\
s06 & 00031097004$^2$ &  2008-01-28& 00:57:00  &  3787.7&  0.584 \\
s07 & 00031097006$^2$ &  2008-01-30& 17:18:01  &  3876.7&  0.803 \\
s08 & 00031097007$^2$ &  2008-01-31& 06:17:01  &  2585.1&  1.518 \\
s09 & 00031097008$^2$ &  2008-02-01& 00:04:00  & 26394.8&  0.182 \\
s10 & 00031097009$^2$ &  2008-02-02& 23:59:01  & 22568.9&  0.999 \\
s11 & 00031097010$^2$ &  2008-02-05& 16:40:01  &  3468.7&  1.396 \\
s12 & 00031097011$^2$ &  2008-02-06& 00:27:26  &  3332.3&  3.913 \\
s13 & 00031097012$^2$ &  2008-02-07& 00:34:46  &  3372.7&  1.334 \\
s14 & 00031097013$^2$ &  2008-02-09& 21:33:01  &  2785.1&  0.923 \\
s15 & 00031097016$^2$ &  2008-02-11& 17:02:01  &  2478.9&  2.168 \\
s16 & 00031097017$^2$ &  2008-02-12& 00:45:01  &  2967.9&  1.300 \\
s17 & 00031097018$^2$ &  2008-02-13& 07:15:01  &  2598.9&  1.590 \\
s18 & 00031097019$^2$ &  2008-02-14& 15:22:01  &  3382.0&  0.412 \\
s19 & 00031097020$^2$ &  2008-02-15& 12:29:01  &  2422.4&  1.264 \\
s20 & 00031097022$^2$ &  2008-02-17& 12:52:01  &  3692.9&  3.464 \\
s21 & 00031097023$^2$ &  2008-02-18& 17:30:00  &  3389.9&  1.234 \\
s22 & 00031097025$^2$ &  2008-02-20& 06:15:01  &  3392.9&  2.647 \\
s23 & 00031097026$^2$ &  2008-02-21& 07:57:01  &  3646.4&  2.027 \\
s24 & 00031097028$^2$ &  2008-02-23& 14:33:01  &  2977.3&  1.415 \\
s25 & 00031097029$^2$ &  2008-02-24& 16:14:00  &  2349.4&  0.638 \\
s26 & 00031097030$^2$ &  2008-02-25& 11:35:01  &  2408.8&  0.816 \\
s27 & 00031097031$^2$ &  2008-02-26& 05:10:01  &  3835.2&  2.159 \\
s28 & 00031097034$^2$ &  2008-03-16& 08:35:01  &  9246.5&  1.688 \\
s29 & 00031308001$^3$ &  2008-12-14& 03:17:01  &  6758.9&  6.929 \\
s30 & 00090033001$^4$ &  2009-03-18& 08:44:01  & 14363.4&  6.457 \\
s31 & 00031097035$^2$ &  2009-04-01& 12:10:22  &  9225.9&  1.270 \\
s32 & 00031097036$^2$ &  2009-04-02& 08:59:01  & 12517.4&  1.741 \\
s33 & 00031097037$^2$ &  2009-04-03& 09:10:01  & 10208.9&  1.064 \\
s34 & 00031097038$^2$ &  2009-04-07& 09:24:01  &  8674.8&  0.911 \\
s35 & 00031097039$^2$ &  2009-04-08& 07:53:01  &  8259.7&  1.191 \\
s36 & 00031097040$^2$ &  2009-04-09& 13:04:00  &  9845.3&  1.465 \\
s37 & 00031097041$^2$ &  2009-06-03& 01:46:01  &  3790.9&  0.348 \\
s38 & 00031097042$^2$ &  2009 06 04& 06:51:00  &  3923.2&  1.858 \\
s39 & 00031097043$^2$ &  2009-06-05& 06:58:00  &  3364.8&  1.595 \\
s40 & 00031097044$^2$ &  2009-06-06& 06:50:01  &  4194.7&  3.771 \\
\hline
\end{tabular}
\\
{Proposers of observations were  $^1$F. Senziani, $^2$A. 
Pollock, $^3$E. Pian  and $^4$M. Corcoran. }
\end{table}

\subsection{ASAS V-band observations}
 WR 25 was observed from 2000 December 3 to 2009 December 3 by 
ASAS\footnote{http://www.astrouw.edu.pl/asas}. We have used only `A' and `B' 
grade data  within  1 arcsec to the target of WR 25.  ASAS photometry  provides 
five sets of  magnitudes corresponding to five aperture values varying in size 
from 2 to 6 pixels in diameter. For bright objects, \cite{pojmanski02} suggested 
that the magnitudes corresponding to the largest aperture (i.e. the aperture 
diameter of 6 pixels) are useful. Therefore, we took  
magnitudes corresponding to the largest aperture for further analysis.

\begin{figure*}
\center
\hspace{-0.5cm}
\subfigure[\xmm -MOS]{\includegraphics[width=8cm]{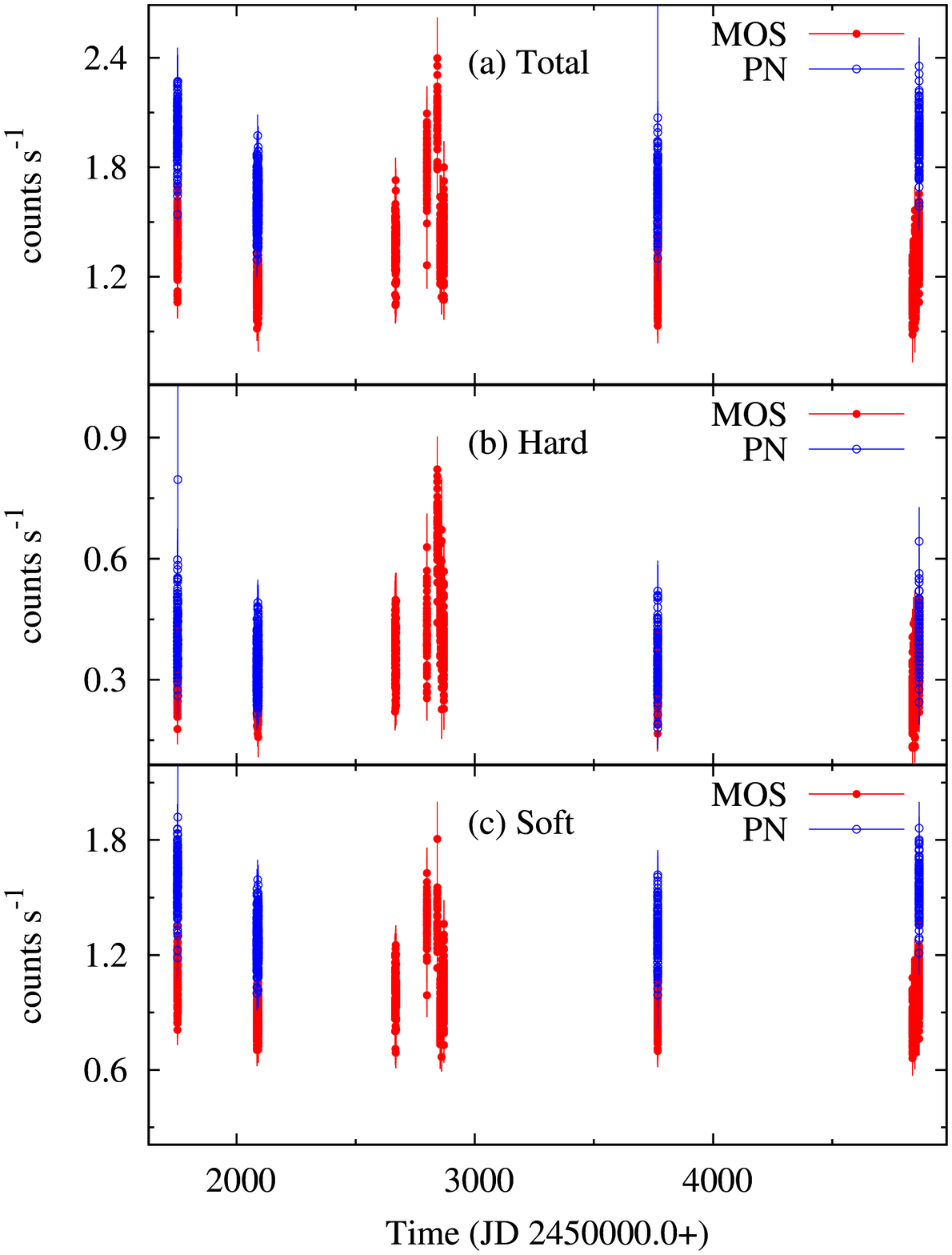}}
\hspace{0.5cm}
\subfigure[\swift -XRT]{\includegraphics[width=8cm]{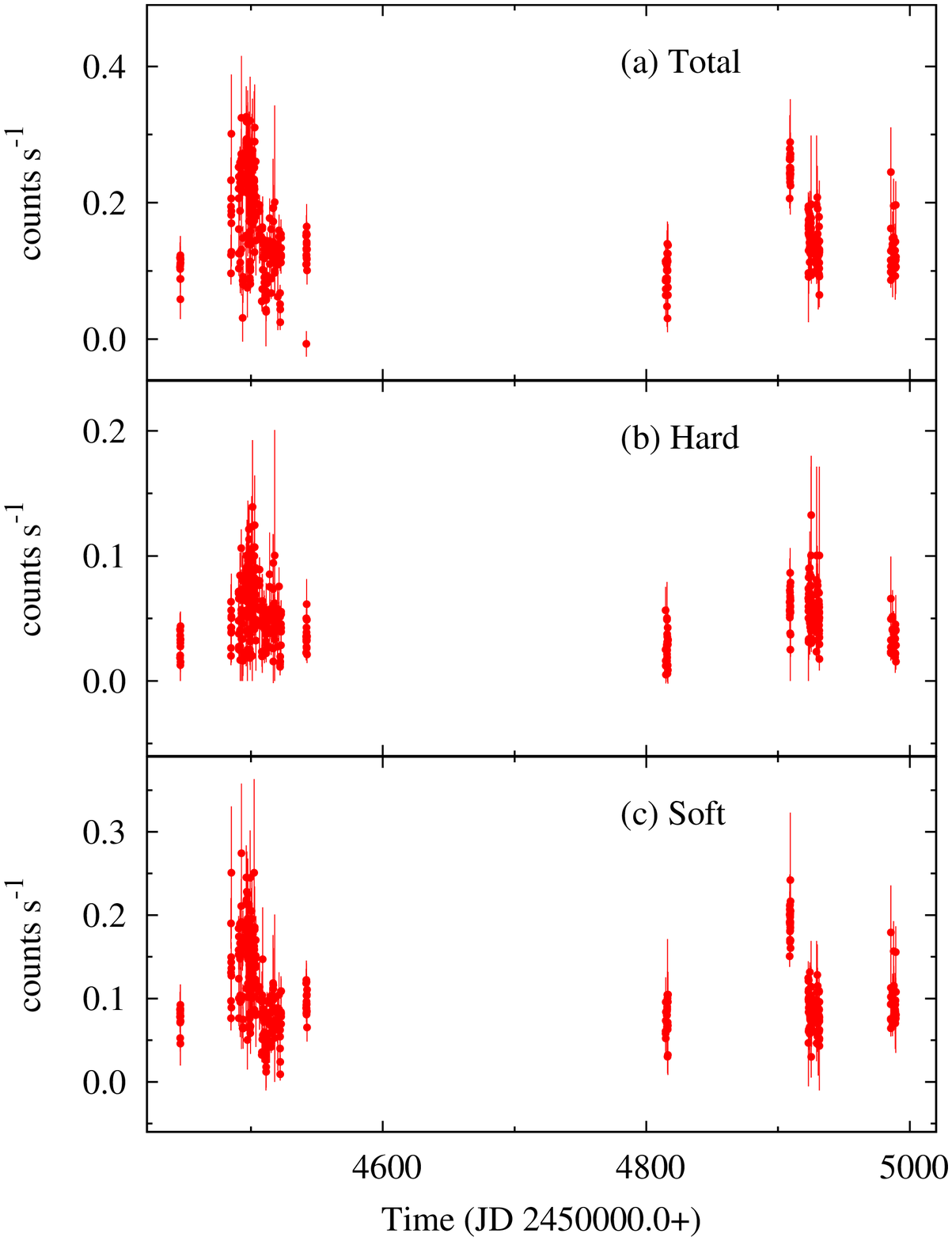}}
\caption{X-ray light curves in total (0.3-10.0 keV), hard (2.0-10.0 keV) and 
soft (0.3-2.0 keV) energy bands as observed from (a) \xmm -EPIC and (b) \swift 
-XRT.} \label{fig:lc_ft}
\end{figure*}

\section{X-ray light curves and period analysis}
\label{sec:lightcurve}
The background-subtracted X-ray light curves as observed from \xmm -EPIC,  and 
\swift -XRT  in the total (0.5-10.0 keV), hard (2.0-10.0 keV) and soft (0.5-2.0 
keV) energy bands are shown in Figures \ref{fig:lc_ft} (a) and  (b), 
respectively running from top to bottom. The light curves were binned at 
2000 s intervals. The variability in each band is clearly seen. For additional confirmation  of 
the variability in all bands, the significance of deviations from the mean count 
rate were measured using the $\chi^2$ - test, defined as

\begin{equation}
\chi^2 = \sum_{i=1}^{N}{\frac{(C_i-\overline{C})^2}{\sigma_i^2}}
\label{eq:chi}
\end{equation}

\noindent where $\overline{C}$ is the average count rate, C$_i$ is the count rate of 
$ith$ observations and $\sigma_i$ is the error corresponding to  C$_i$. The  
$\chi^2$ statistic was compared against a critical value  ($\chi^2_\nu$) for 
99.9\%  significance level, obtained from the  $\chi^2$-probability function. 
For the \xmm ~MOS light curves, $\chi^2$ values were obtained to be 6440, 4403 
and 4499 in the total, hard and soft bands, respectively. 
 These values of $\chi^2$ are very large in comparison to the $\chi_{\nu}^2$ of 1430 for 1599 degrees of freedom.
For  \swift ~-XRT light curves in total, hard and soft 
bands,   $\chi^2$  values of  5188, 5113 and 5124 were obtained, 
respectively, with  $\chi^2_\nu$ $=$ 393 for 483 degrees of freedom. 
This indicates that  WR 25 is essentially variable in all X-ray bands.

\begin{figure}
\includegraphics[width=8.5cm]{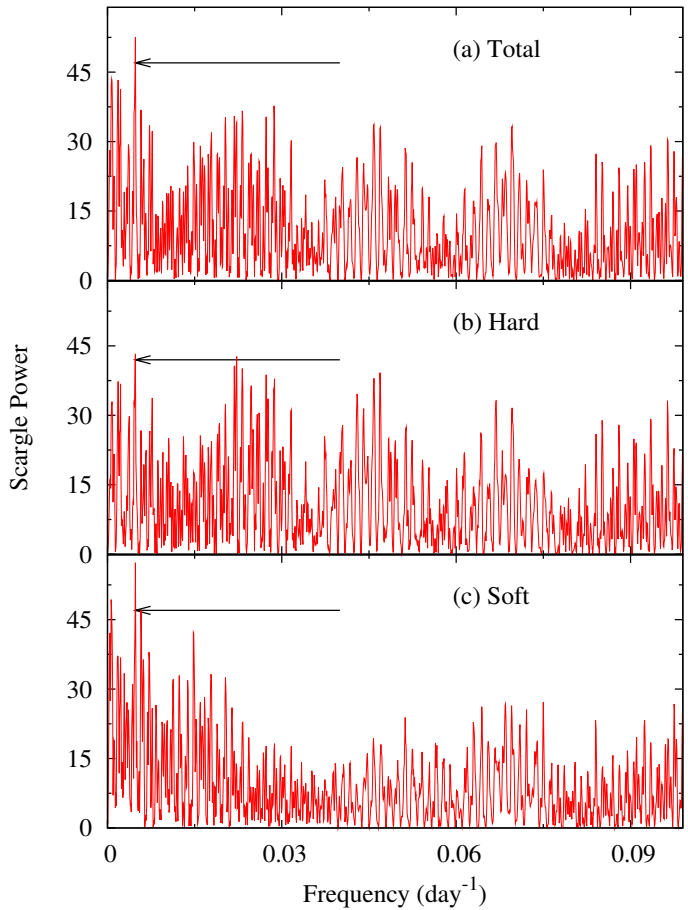}
\caption{Lomb-Scargle power spectra of X-ray data in different energy bands as 
observed from EPIC detectors of \xmm.}
\label{fig:ft}
\end{figure}

The present long-term X-ray data as observed from \xmm ~permit us to derive the 
orbital period of WR 25. We have performed a period analysis of  light curves in 
all bands with a Lomb-Scargle periodogram \citep{lomb76,scargle82}. The plots 
from top to bottom in Figure\ref{fig:ft} show the power spectra  in the total, 
hard and soft bands, respectively. The frequency corresponding to the highest 
peak was found to be  $0.00482\pm0.00008$ cycles day$^{-1}$ in all bands. Other 
frequencies were also noticed in the power spectra but are not consistent with 
each other among all the energy bands. Thus, the corresponding period of 
$207.5\pm3.4$ days appears to be  real. Also, the derived period is very 
similar to the orbital period (=$207.85\pm0.02$ days) derived by \citet{gamen06} 
using radial velocity measurements.  

\begin{figure*}
\hspace{-0.3cm}
\subfigure[XMM-MOS]{\includegraphics[width=8cm]{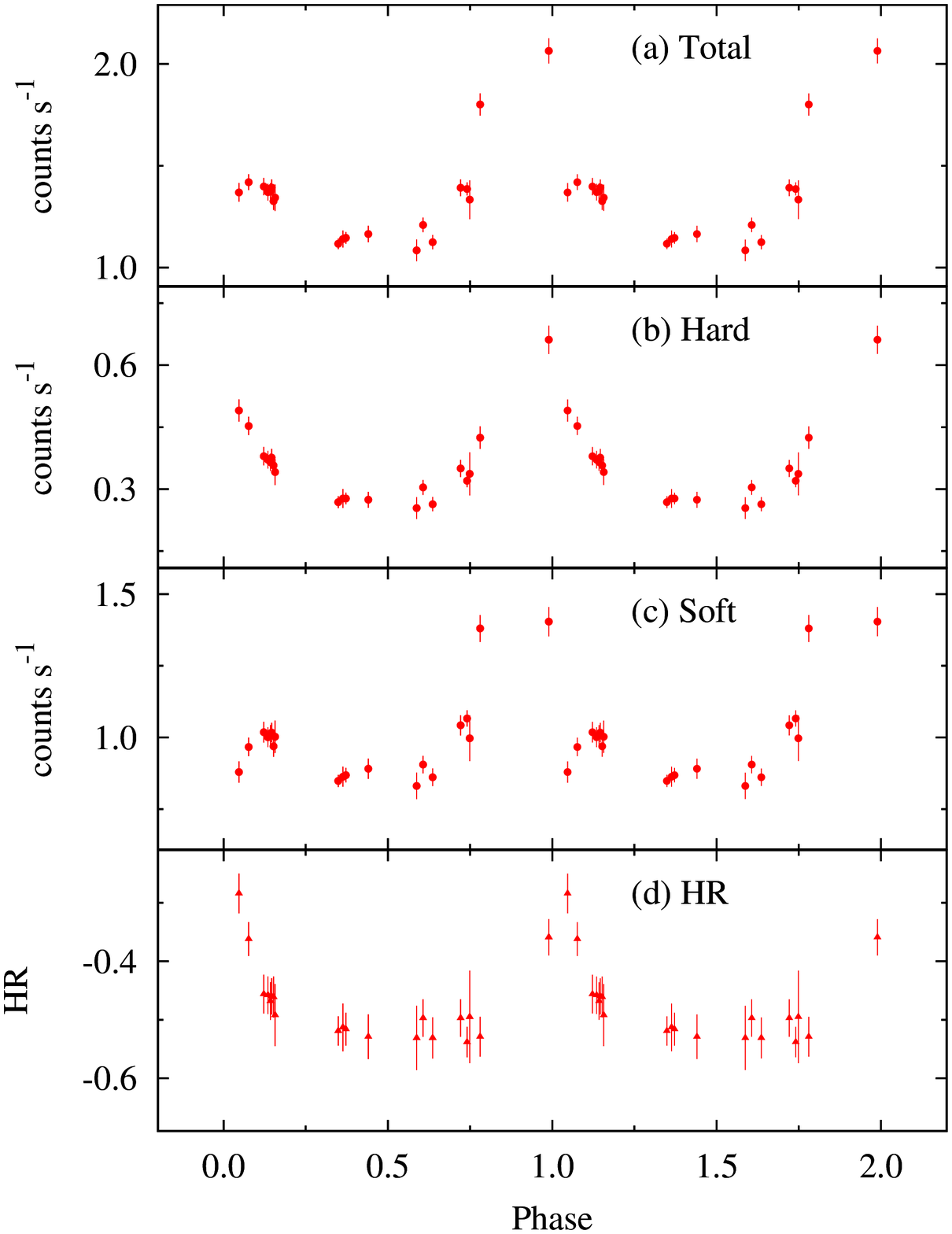}}
\hspace{1.5cm}
\subfigure[SWIFT-XRT]{\includegraphics[width=8cm]{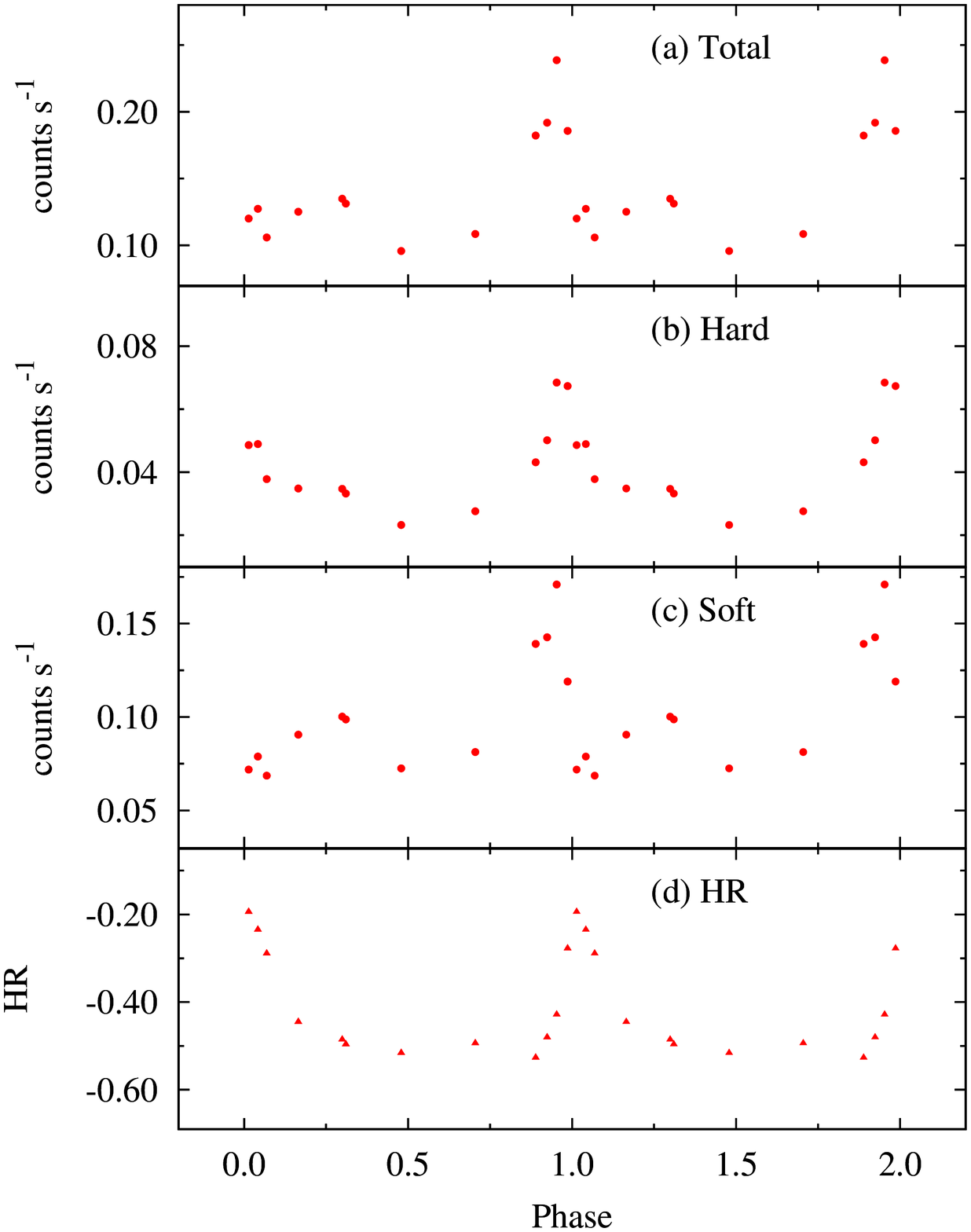}}
\caption{Folded light curves using the ephemeris derived by \cite{gamen06} in 
different energy bands as observed from (a) \xmm ~(b) \swift.}
\label{fig:flc}
\end{figure*}

Further, the X-ray light curves as observed from \xmm-MOS and \swift-XRT in the 
total, hard and soft energy bands were folded using the ephemeris HJD = 
2451958.0 + 207.85E \citep{gamen06} and are shown in Figures \ref{fig:flc}(a) and  (b), respectively. The zero phase in the folded light curves 
corresponds to the time of periastron (HJD 2451958.0). X-ray light curves in 
each band from both observations show similar behavior. The light curves in the 
individual bands show  phase-locked variability. In the soft and total-band 
light curves, the count rates decrease when going from orbital phase 0.0 
(i.e., near the periastron passage) to orbital phase  $\sim 0.03$ then 
increase up to  phase $\sim 0.2$. Count rates further decrease to  phase 
$\sim 0.7$ before reaching  a maximum value near   periastron passage.  
However, in the hard band, the count rates were systematically  decreasing when going 
from periastron to apastron. The ratio of maximum to minimum count rates in 
total, soft and hard bands were found to be 1.9, 2.6, and 1.7 for \xmm-MOS 
observations, and 2.5, 2.6, and 2.9 for \swift-XRT observations, respectively.  

The hardness ratio (HR) defined as (Hard -Soft)/(Hard+Soft) can reveal 
information about the spectral variations.   The HR curve displayed in  plot (d) 
of Figure \ref{fig:flc} exhibits similar behavior to that of the light curves in 
the hard band. The maximum value of the HR during   periastron passage indicates 
a harder spectrum.

\section{X-ray Spectral Analysis}
\label{sec:spectra} 
\begin{figure*}
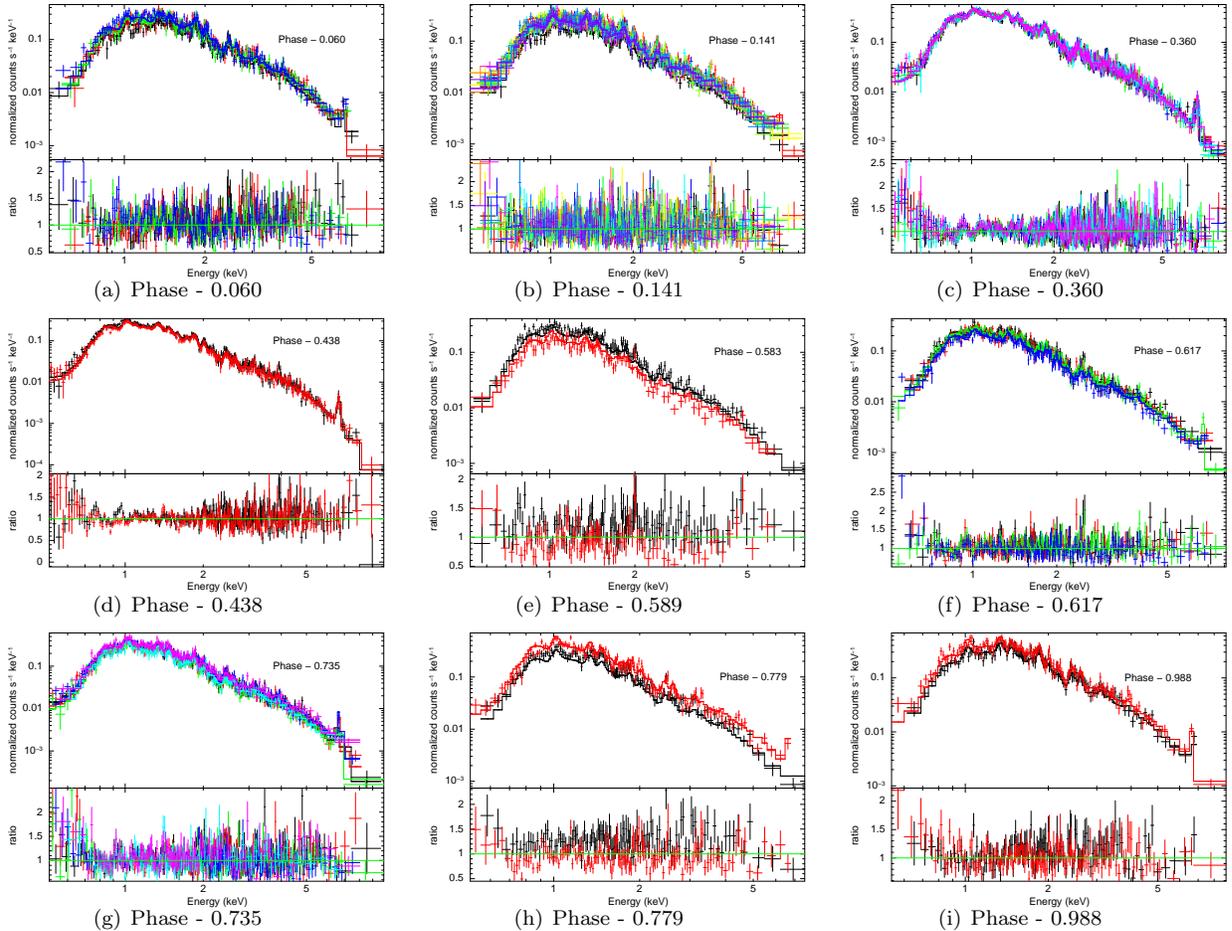

\center
\subfigure[Phase - 0.060]{\includegraphics[angle=-90,width=5.5cm]{fig4a.ps}}
\subfigure[Phase - 0.141]{\includegraphics[angle=-90,width=5.5cm]{fig4b.ps}}
\subfigure[Phase - 0.360]{\includegraphics[angle=-90,width=5.5cm]{fig4c.ps}}
\subfigure[Phase - 0.438]{\includegraphics[angle=-90,width=5.5cm]{fig4d.ps}}
\subfigure[Phase - 0.589]{\includegraphics[angle=-90,width=5.5cm]{fig4e.ps}}
\subfigure[Phase - 0.617]{\includegraphics[angle=-90,width=5.5cm]{fig4f.ps}}
\subfigure[Phase - 0.735]{\includegraphics[angle=-90,width=5.5cm]{fig4g.ps}}
\subfigure[Phase - 0.779]{\includegraphics[angle=-90,width=5.5cm]{fig4h.ps}}
\subfigure[Phase - 0.988]{\includegraphics[angle=-90,width=5.5cm]{fig4i.ps}}
\caption{X-ray spectra of WR 25  as observed from MOS and PN detectors  along 
with the best fit folded 2T {\sc vapec} model at nine epochs.   WR 25 was 
observed with the PN detector only at three phases, namely 0.360, 0.438 and 
0.735. The PN spectra are shown on the top at these three epochs. The lower 
panels show the residual in terms of the ratio of the data and model. The 
emission lines of NeX, Mg XI+XII, Si III+IV, S XV+XVI, Ar XVII+XVIII, Ca XIX+XX 
and Fe (K-shell) are seen in the spectra.}
\label{fig:spec}
\end{figure*}

X-ray spectra of WR 25 as observed by  the EPIC detector of \xmm 
~at different orbital phases are shown in Figure \ref{fig:spec}. Below 1 keV, 
the  X-ray spectra were found to be affected by  high extinction. Strong 
emission lines like  Fe XVII (0.8 keV), Ne X (1.02 keV), Mg XII (1.47 keV), Si 
XIII (1.853 keV), S XV (2.45 keV), Ar XVII (3.12 keV),Ca XIX+XX (3.9 keV), and 
Fe XXV (6.67 keV) were identified in the X-ray spectra of WR 25.   
In order to
derive the spectral parameters at different orbital phases, we performed 
spectral analysis of each data set  corresponding to different orbital phases  
using simultaneous/joint fitting of EPIC data by models of the Astrophysical 
Plasma Emission Code [APEC; \cite{smith01}], as implemented in the X-ray spectral 
fitting package {\sc xspec} \citep{arnaud96} version 12.7.1. For spectral fitting,  
we adopted the similar approach of  \citet{raassen03} and 
\citet{pollock06}. The form of the model used was 
{\sc wabs(ism)*wabs(local)*(vapec+vapec)}. A $\chi^2$ minimization gave the best
fit model to the data. The presence of interstellar material along the line of 
sight and the local circumstellar material around the stars can modify the X-ray 
emission from massive stars. We have applied  the local absorption in the line of 
sight to the source using photoelectric absorption cross-sections according to  
\cite{morrison83} and modeled as {\sc wabs} with two absorption components i.e. 
interstellar medium (\nhism) and local (\nhl) hydrogen column densities.
Assuming the  E(B-V ) $\sim$ 0.63  mag \citep{vanderhucht01} and  a normal 
interstellar reddening law towards WR 25, and using the relation given by 
\cite{gorenstein75}, the \nhism ~ was estimated to be $\sim 3.7\times10^{21}$ 
cm$^{-2}$.  For the first stage of  spectral fitting, we fixed the \nhism ~and 
abundances of He(=2.27), C(=0.15), and N(=5.9),  and varied other parameters for 
all phases. The values of He, C, and N abundances were adopted from the optical 
spectrum of WR 25 \citep{crowther95}. The temperatures for both components  were 
found to be constant within the $1\sigma$ level at all phases.  However, abundances 
of Ne, Mg, Al, Si, Ca, Ar, Fe, and Ni were found to be constant within the 1-2 
$\sigma$ level.  Average values of elemental abundances and temperatures  are 
given in Table \ref{tab:abun}.  In the next stage of spectral fitting, we fixed  
temperatures and abundances of all elements at their average values for all 
phases along with \nhism, and varied \nhl , and normalizations of both components. 
Many spectra were taken at the same orbital phase; therefore, joint spectral 
fittings were performed for those spectra, which were observed within a difference
of $\leq 0.02$ in the orbital phase. 
In this way, from \xmm ~observations, we have 
nine data points over an orbital cycle of WR 25. The final set of the 
best-fitted parameters is given in Table \ref{tab:spec}.

The  method used for the fitting of \xmm-EPIC spectra  was also applied for the fitting of 
spectra observed from the \swift-XRT. Since WR 25 was observed more than 40 
times by the  \swift-XRT and many observations were taken with short exposures, 
the spectra with a phase difference of $\leq$0.02 were fitted jointly. 
In this way, we have spectra at 12 different phases over an orbital cycle from 
\swift ~observations. For the  spectral fitting of \swift-XRT data, we have 
fixed temperatures and  abundances to values as obtained from spectral fitting 
of \xmm-EPIC data. However, \nhl ~and normalizations of both  components were 
kept as free parameters in the spectral fitting. The best fit parameters from 
the spectral fitting of \swift~-XRT data  are also given Table \ref{tab:spec}. 
We have compared  various parameters derived from both observatories.  It appears 
that the X-ray fluxes and other spectral parameters derived from spectral fitting of 
both satellite data are consistent. Comparison of spectral parameters derived 
from both telescopes may not be possible  due to the lack of simultaneous 
observations. However, \cite{plucinsky12} has shown that the spectral parameters
 from both observations are in agreement within $\pm10\%$.

\begin{figure*}
\subfigure[XMM-Newton EPIC observations]{\includegraphics[width=8cm]{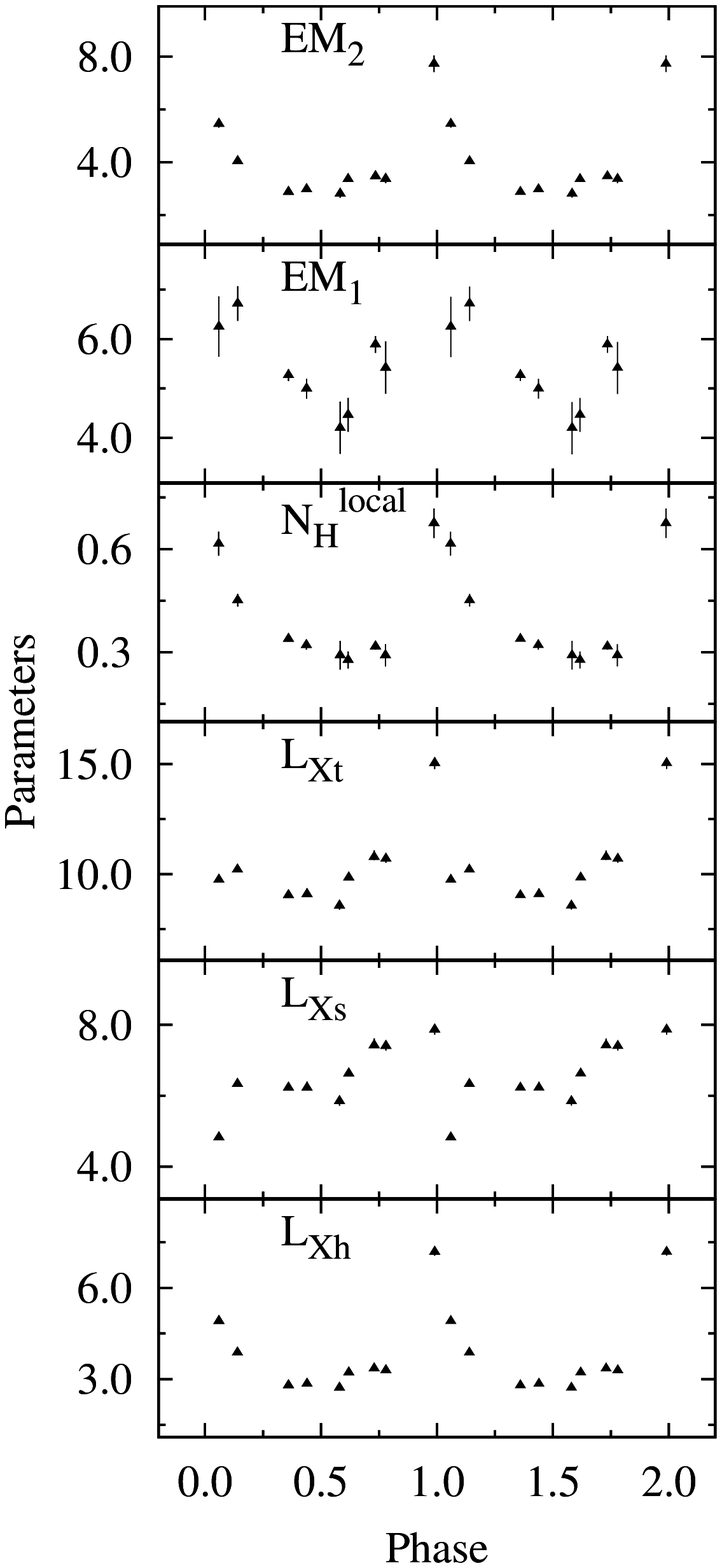}}
\subfigure[Swift XRT observations]{\includegraphics[width=8cm]{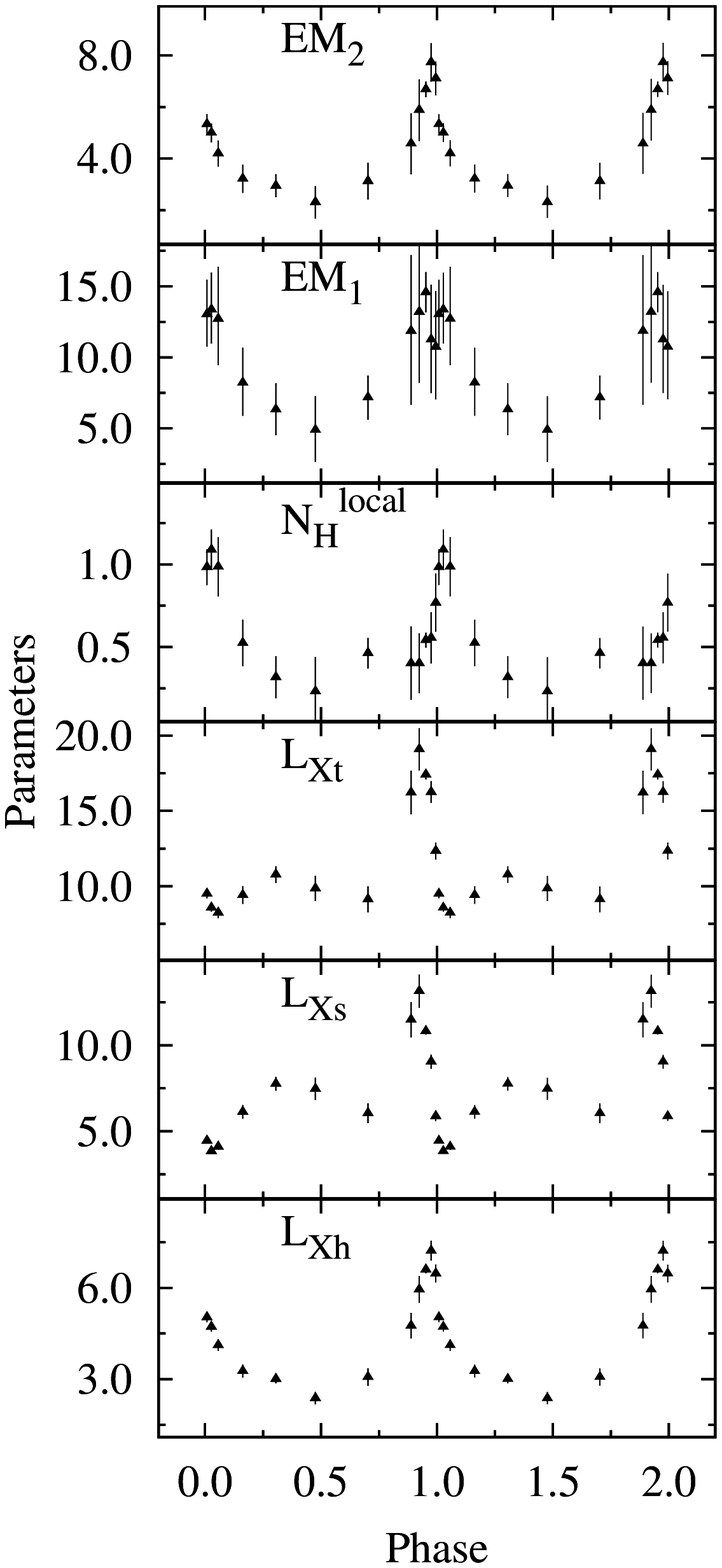}}
\caption{Spectral parameters as a function of orbital phase as observed from  
(a) \xmm -EPIC and (b) \swift -XRT, where X-ray luminosities are in  units of 
$10^{33}$ \lum, \nhl ~is in units of $10^{22}$ cm$^{-2}$, and \emc ~and \emh 
~are in  units of 10$^{56}$ cm$^{-3}$. }
\label{fig:para}
\end{figure*}

The X-ray fluxes of WR 25  are estimated using the {\sc cflux} model in {\sc 
xspec} and are corrected for \nhism. 
The  X-ray luminosities of WR 25 in soft 
(\lxs), hard (\lxh) and total (\lxt) energy bands were derived by using the 
corresponding unabsorbed flux values and a distance of 3.24 kpc. The EMs (EMs), 
\emc ~and \emh, ~corresponding to both the cool and the hot plasma components are derived 
from the normalization parameters.  We have plotted \lxs, \lxh, \lxt, \nhl, \emc,
and \emh ~as a function of orbital phase in Figure \ref{fig:para} (a) and 
\ref{fig:para} (b) for observations from \xmm ~and \swift, respectively. The 
maximum value of \lxs ~was  found near  periastron passage and dropped suddenly 
to a phase $\sim 0.03$ and then increased to  phase $\sim 0.3$. After phase 
$\sim 0.3$,  \lxs ~decreased to  phase $\sim 0.7$. It appears that \lxs 
~peaked twice during an orbital cycle. The minimum value \lxs ~was  observed
just after  periastron passage, which was $\sim 3.2$ times lower than the 
observed maximum value. A similar tendency of X-ray luminosity  was also observed 
in the total energy band, where the maximum to minimum flux ratio was found to be $\sim 2.3$. 
Apart from the periastron passage, no other peak of \lxh ~was noticed in an 
orbital cycle of WR 25. The minimum value of \lxh ~was observed near  phase 
$\sim 0.5$ and was $\sim 3.1$ times lower than that observed during the 
periastron passage. The value of \nhl ~was observed to be maximum after the periastron passage 
at phase 0.03. However, the minimum value of \nhl ~was observed at phase 
$\sim 0.5$. The EMs, \emc ~and \emh ~corresponding to the cool and 
the hot temperature components  were also found to be phase 
dependent, with maximum at periastron passage and minimum at  apastron passage.

\begin{table}
\caption{Average values of temperatures and abundances of WR 25 as obtained from 
X-ray spectral fitting. Abundances are in units of Solar photospheric values.}
\label{tab:abun}
\begin{tabular}{lcllc}
\hline
\hline
Parameters  & Value &  & Parameters  & Value \\
\hline
kT$_1$ (keV)& 0.628$_{-0.007}^{+0.009}$ & & Si& 1.07$_{-0.07}^{+0.09}$  \\
kT$_2$ (keV)& 2.75$_{-0.05}^{+0.06}$    & & S & 1.59$_{-0.10}^{+0.11}$\\
 O          & $< 0.1$                   & & Ar& 1.30$_{-0.22}^{+0.27}$\\
 Ne         & 1.60$_{-0.14}^{+0.19}$    & & Ca& 1.67$_{-0.38}^{+0.42}$\\
 Mg         & 1.38$_{-0.09}^{+0.12}$    & & Fe& 0.68$_{-0.04}^{+0.05}$\\
 Al         & 4.42$_{-0.86}^{+1.28}$    & & Ni& 3.93$_{-0.68}^{+0.81}$  \\
\hline
\end{tabular}
\end{table}


\section{Optical V-band light curve}
\label{sec:optical}
\begin{figure}
\subfigure[V-band light curve]{\includegraphics[width=6cm,angle=-90]{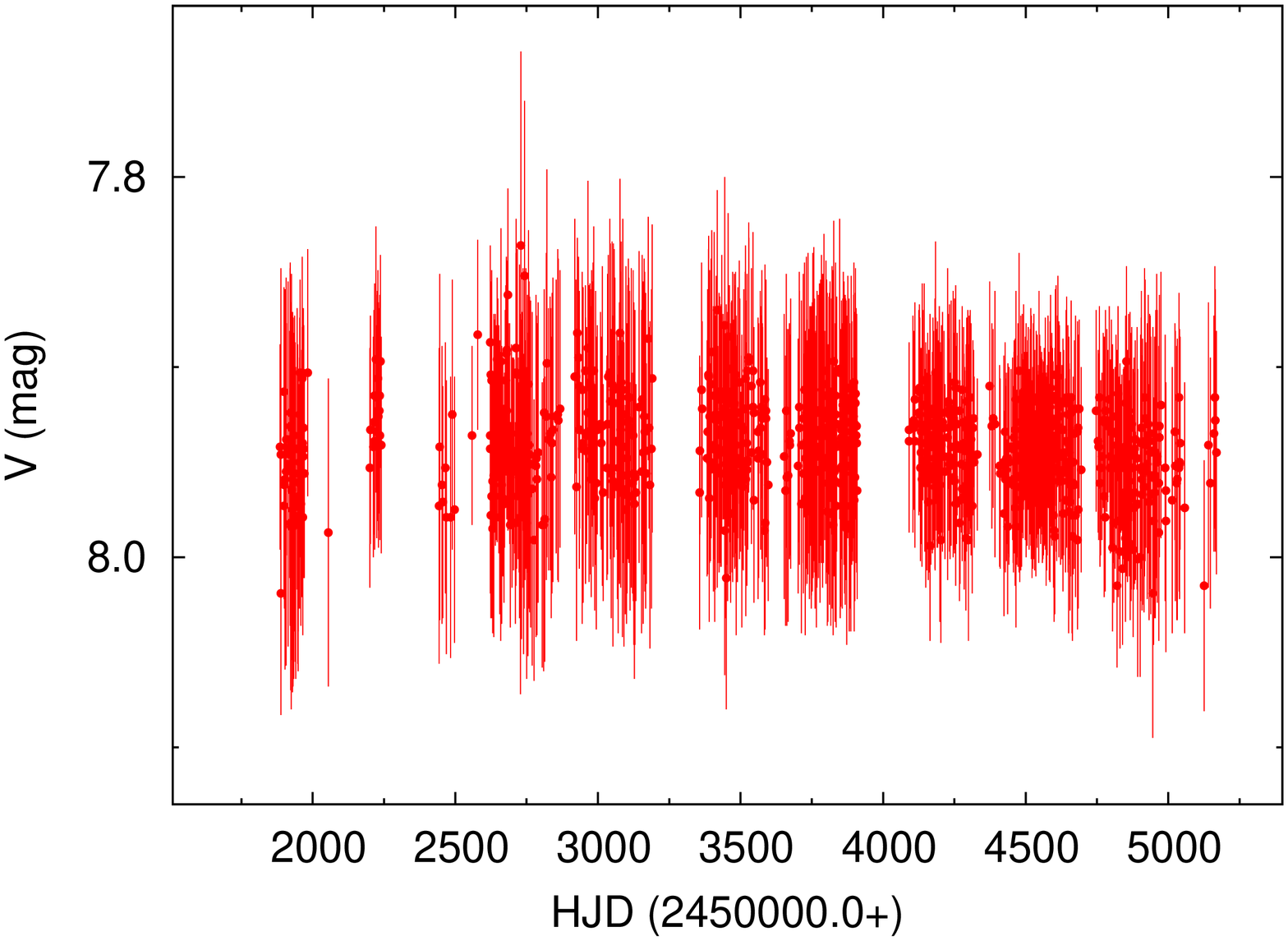}}
\subfigure[V-band light curve (running average)]{\includegraphics[width=6cm,angle=-90]{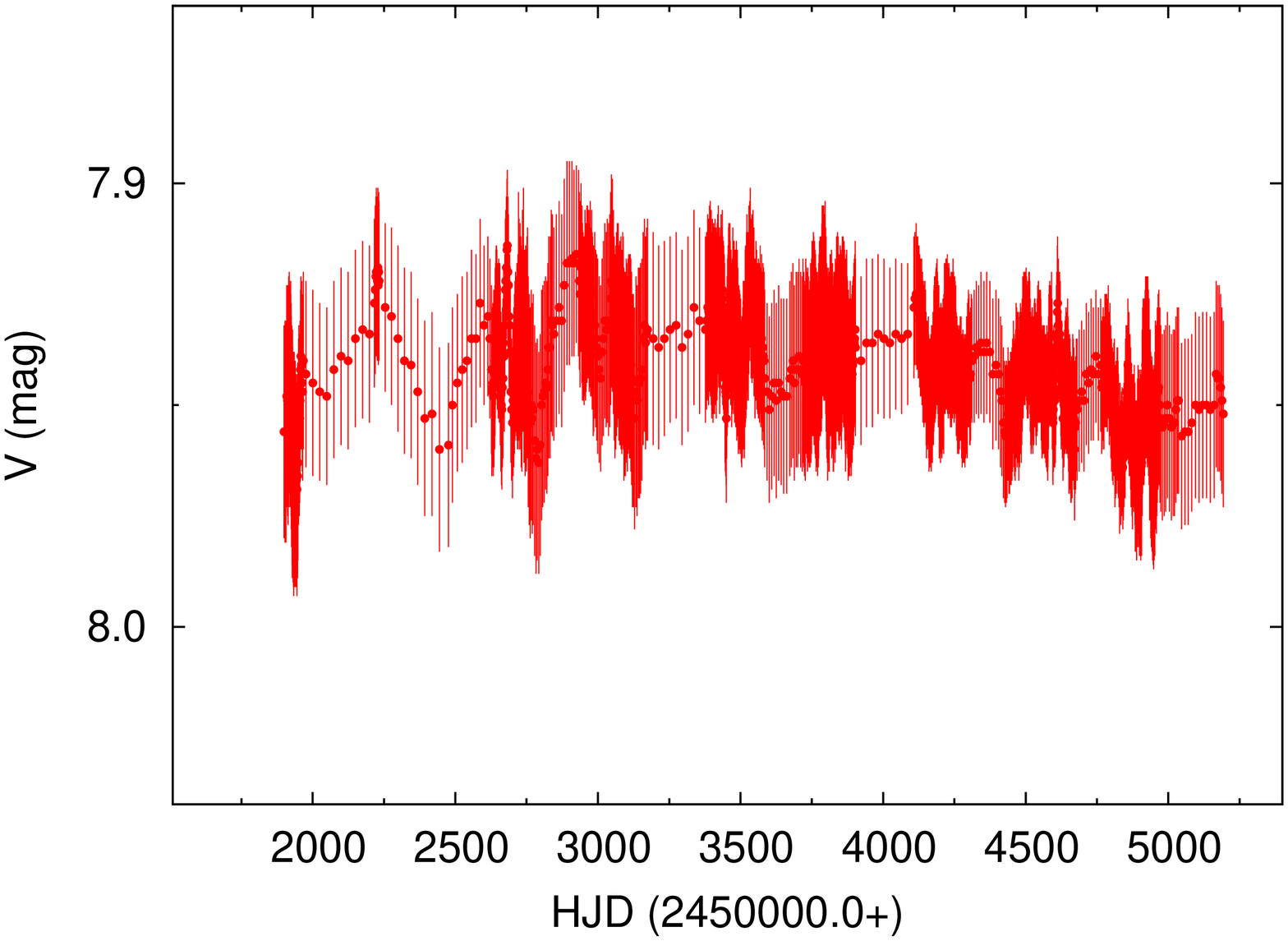}}
\subfigure[Folded light curve]{\includegraphics[width=5cm,angle=-90]{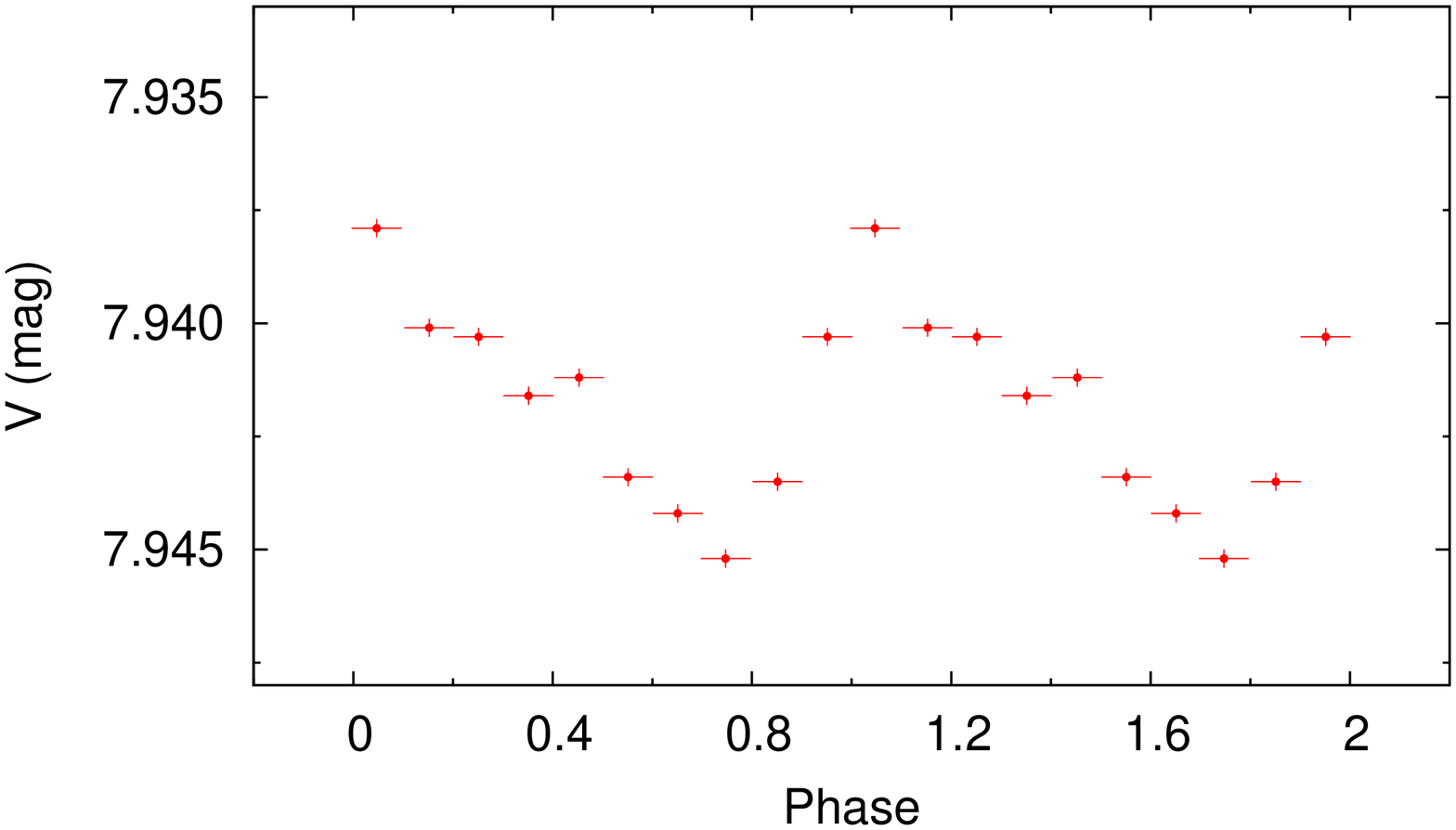}}
\caption{(a) V-band light curve WR 25 as observed in ASAS, (b)  light curve 
after smoothing (see the text for details) (c) folded light curve of WR 25 using the 
ephemeris  HJD = 2451598.0 + 207.85 E as derived by \cite{gamen06}.}
\label{fig:asas}
\end{figure}

The V-band light curve of WR 25 as obtained from the ASAS archive is plotted in 
Figure \ref{fig:asas} (a).   In ASAS observations, error bars are very large (i.e. 
0.035 to 0.1 mag). Therefore, it is difficult  to search for any small scale 
variability that might be present. 
In order to reduce the short term 
fluctuations, a moving average of 10 data points in the forward direction was 
performed. The light curve of the moving-averaged  data points is shown in Figure 
\ref{fig:asas} (b). The V-band light curve does not appear to be constant over 
the time span of  the observations; however, variations are not statistically 
significant. The  $\chi^2$ was  found to be 270 for 956 degrees of freedom 
(see Equation \ref{eq:chi}). This value of  $\chi^2$ is very low in 
comparison to the $\chi^2_\nu$ of 826 corresponding to the 99.9\% significance 
level.  Furthermore,  we have folded the V-band data with phase bins of 0.1 and using the ephemeris given by \cite{gamen06}. The folded V-band light curve of WR 25 is shown in
Fig \ref{fig:asas} (c).  It appears that variability is present in the 
light curve. The variability amplitude was found to be $7.3\pm0.2$
mmag. The maximum brightness was seen near  periastron passage and after 
that the brightness decreased toward apastron passage,  being minimum near
phase $\sim$0.75.


\section{Discussion and conclusions}
\label{sec:discussion}

We have carried out X-ray and optical  studies of WR 25 using  X-ray data  
from the \xmm ~and \swift satellites and optical data from the ASAS archive.  The 
X-ray spectra of WR 25 at all orbital phases are well-explained by a 
two-temperature  plasma model. The temperatures of both components were found to be 
constant throughout an orbital cycle. The temperature values are similar 
to those derived by \cite{pollock06} and \cite{raassen03} using \xmm ~data.  The 
temperature of the cool component (=0.628 keV) of WR 25 is comparable to
that of other WN-subtype WR+O binaries: for example WR1 [0.56-0.67 keV;  
\citep{ignace03}], WR 22 [0.6 keV; \citep{gosset09}] WR 47 [$0.57\pm0.05$ keV; 
\citep{bhatt10a}], WR 139 [$0.63\pm0.05$ \citep{bhatt10a}], and  WR 147
[0.7-0.8 keV; \citep{skinner07}]. A plasma temperature of 0.6-0.8 keV is also 
dominant for massive OB stars \citep{sana06,naze09,bhatt10b}.  The soft X-ray 
component may originate in the winds of the individual components of WR 25 via 
radiation driven instability shocks.  The ratio between the wind momentum 
($\dot{M} v_\infty$) and the momentum of the radiation field ($L/c$) for WR 25 
was derived to be $\sim 1$ \citep{hamann06}, indicating that  wind of WR 25 is 
driven by radiation pressure. Furthermore, the derived  temperature of the 
cool component appears to be realistic for radiation-driven wind shocks.  The 
derived plasma temperature for WR 25 can be used to estimate the  pre-shock velocity 
using the relation $kT_{sh} = 1.95\mu v_{1000}^2$ keV \citep{luo90}, where $\mu$ 
is mean mass per particle in units of the proton's mass (1.16 for a WN star and 0.62 
for an O-type star), and $v_{1000}$ is shock velocity in units of 1000 km s$^{-1}$. 
The pre-shock velocities corresponding to the cool temperature for the WN and O 
components of WR 25 were found to be  527 and 721 km s$^{-1}$, respectively.  
These values are about a factor of $\sim 2$ more than those predicted by the 
radiative shock model of \cite{lucy82}. The advanced version of the wind-shock
model by \cite{owocki88} predicts X-ray emission up to 1 keV.  The hydrodynamic 
shocks, which are expected to occur within unstable stellar winds of massive 
O-type  stars, also show  similar pre-shock velocities \citep{feldmeier97}.  
Observationally some apparently single WN stars (e.g. WR 1, WR 6,  and WR 110) show 
intrinsic X-ray emission \citep{skinner02a,skinner02b,ignace03} while apparently single WC stars  and the WN-subtype WR star WR 40  have not been detected in X-rays thus far \citep{oskinova05, gosset05}.

The temperature corresponding to hot the plasma of WR 25 is  intermediate compared to those 
observed for other similar WR binaries \citep{zhekov12}.  Using  the mean particle 
weight for WN stars, the maximum shock temperature on the line of centers  for an 
adiabatic shock corresponds to a pre-shock wind velocity of $1103 \pm 12$ km 
s$^{-1}$ (see Luo et al. 1990), which  is $\sim 50$ \% of the observed terminal 
velocity \citep{crowther98,niedzielski02} of WR 25.   This could be due to 
an oblique wind collision over  most parts of the shock, which occurs with a 
lower velocity normal to the shock and thus leads to a lower plasma temperature.
The derived values of \nhl ~and X-ray luminosities exhibit  phase-related time 
variability. We found \nhl ~to be maximum after  periastron passage and 
minimum when \lxt ~is maximum during  periastron passage. It also appears 
that \nhl ~reached a  peak value during the eclipse, indicating that the extra absorption 
could be due to winds from the WN star. The column density along the line of 
sight through the WR wind to the region of stellar wind collision is \nhl $\propto 
\dot{M} D^{-1} v_\infty^{-1}$ \citep{usov92}, where $D$ is the distance between 
the binary components. Using the parameters of WR 25, \nhl ~was estimated to be 
$\sim 1.36\times10^{22}$ cm$^{-2}$, which is quite similar to the observed  
maximum value (see Table \ref{tab:spec}).  

\begin{table*}
\begin{center}
\caption{Best Fit Parameters Obtained From X-ray Spectral Fitting of WR 25.}
\label{tab:spec}
\begin{tabular}{lcccccccccccc}
\hline
\hline
Set          &  Phase  & \nhl                  & \emc                  &\emh                  & \lxs                  &\lxh                  &\lxt                   &$\chi^2_\nu$(dof)\\
\hline
&&&&XMM\\
\hline
x13,x14      &  0.060  & 0.62$_{-0.03}^{+0.03}$& 6.25$_{-0.61}^{+0.61}$&5.45$_{-0.14}^{+0.14}$& 4.82$_{-0.06}^{+0.06}$&4.91$_{-0.06}^{+0.06}$& 9.74$_{-0.12}^{+0.12}$&1.18( 598) \\
x06,x07,x08, &  0.141  & 0.45$_{-0.02}^{+0.01}$& 6.72$_{-0.35}^{+0.35}$&4.03$_{-0.09}^{+0.09}$& 6.33$_{-0.06}^{+0.06}$&3.87$_{-0.03}^{+0.03}$&10.20$_{-0.09}^{+0.09}$&1.07(1048) \\
x09,x10,x15  &         &                       &                       &                      &                       &                      &                       &           \\
x03,x04,x05  &  0.360  & 0.34$_{-0.01}^{+0.01}$& 5.27$_{-0.12}^{+0.12}$&2.87$_{-0.03}^{+0.03}$& 6.22$_{-0.06}^{+0.09}$&2.80$_{-0.01}^{+0.07}$& 9.03$_{-0.02}^{+0.14}$&1.23(2956) \\
x16          &  0.438  & 0.32$_{-0.01}^{+0.01}$& 5.00$_{-0.20}^{+0.20}$&2.97$_{-0.06}^{+0.06}$& 6.22$_{-0.06}^{+0.17}$&2.85$_{-0.05}^{+0.10}$& 9.08$_{-0.11}^{+0.27}$&1.32(965)  \\
x17          &  0.583  & 0.29$_{-0.04}^{+0.03}$& 4.20$_{-0.52}^{+0.53}$&2.80$_{-0.15}^{+0.15}$& 5.84$_{-0.12}^{+0.12}$&2.72$_{-0.06}^{+0.05}$& 8.56$_{-0.18}^{+0.17}$&1.37(227)  \\
x18,x19      &  0.617  & 0.28$_{-0.02}^{+0.02}$& 4.47$_{-0.34}^{+0.35}$&3.36$_{-0.10}^{+0.10}$& 6.62$_{-0.07}^{+0.08}$&3.22$_{-0.04}^{+0.04}$& 9.84$_{-0.11}^{+0.11}$&1.12( 554) \\
x01,x02,x20  &  0.735  & 0.32$_{-0.01}^{+0.01}$& 5.89$_{-0.17}^{+0.17}$&3.46$_{-0.05}^{+0.05}$& 7.42$_{-0.07}^{+0.19}$&3.35$_{-0.06}^{+0.11}$&10.77$_{-0.13}^{+0.30}$&1.30(2336) \\
x11          &  0.779  & 0.29$_{-0.03}^{+0.03}$& 5.42$_{-0.52}^{+0.53}$&3.36$_{-0.16}^{+0.16}$& 7.40$_{-0.13}^{+0.13}$&3.29$_{-0.06}^{+0.06}$&10.69$_{-0.18}^{+0.18}$&1.28(270)  \\
x12          &  0.988  & 0.68$_{-0.04}^{+0.04}$&12.77$_{-1.36}^{+1.39}$&7.73$_{-0.32}^{+0.31}$& 7.86$_{-0.14}^{+0.14}$&7.18$_{-0.12}^{+0.13}$&15.04$_{-0.26}^{+0.26}$&1.25(281)  \\  
\hline
&&&&Swift\\
\hline
s15,s16,s17,            &0.009                  & 0.98$_{-0.11}^{+0.11}$& 13.04$_{-2.29}^{+2.43}$& 5.34$_{-0.39}^{+0.39}$& 4.45$_{-0.14}^{+0.14}$ & 5.03$_{-0.16}^{+0.16}$&  9.49$_{-0.30}^{+0.30}$&1.40 (178) \\           
s32,s33                 &                       &                       &                        &                       &                        &                       &                        &           \\  
s19,s20,s21,            &0.028                  & 1.09$_{-0.12}^{+0.13}$& 13.37$_{-2.39}^{+2.59}$& 5.01$_{-0.37}^{+0.36}$& 3.85$_{-0.12}^{+0.12}$ & 4.72$_{-0.15}^{+0.15}$&  8.57$_{-0.27}^{+0.27}$&1.21 (191) \\           
s34,s35,s36             &                       &                       &                        &                       &                        &                       &                        &           \\     
s22,s23,s24,            &0.057                  & 0.99$_{-0.18}^{+0.19}$& 12.72$_{-3.27}^{+3.66}$& 4.21$_{-0.51}^{+0.50}$& 4.11$_{-0.19}^{+0.19}$ & 4.12$_{-0.19}^{+0.19}$&  8.23$_{-0.37}^{+0.37}$&1.15 (103) \\           
s25,s26,s27             &                       &                       &                        &                       &                        &                       &                        &           \\     
s28                     &0.163                  & 0.52$_{-0.14}^{+0.14}$& 8.22$_{-2.34}^{+2.46}$ & 3.22$_{-0.55}^{+0.54}$& 6.13$_{-0.39}^{+0.39}$ & 3.27$_{-0.21}^{+0.21}$&  9.41$_{-0.59}^{+0.59}$&1.17 ( 49) \\           
s37,s38,s39,            &0.306                  & 0.32$_{-0.12}^{+0.12}$& 6.34$_{-1.82}^{+1.85}$ & 2.96$_{-0.45}^{+0.44}$& 7.76$_{-0.40}^{+0.40}$ & 3.01$_{-0.15}^{+0.15}$& 10.77$_{-0.55}^{+0.55}$&1.12 ( 85) \\           
s40                     &                       &                       &                        &                       &                        &                       &                        &           \\     
s29                     &0.476                  & 0.23$_{-0.21}^{+0.18}$& 4.90$_{-2.27}^{+2.38}$ & 2.32$_{-0.64}^{+0.61}$& 7.47$_{-0.64}^{+0.64}$ & 2.38$_{-0.20}^{+0.20}$&  9.85$_{-0.84}^{+0.84}$&1.16 ( 31) \\           
s01                     &0.703                  & 0.46$_{-0.09}^{+0.11}$& 7.18$_{-1.55}^{+1.55}$ & 3.13$_{-0.71}^{+0.71}$& 6.05$_{-0.57}^{+0.57}$ & 3.07$_{-0.29}^{+0.29}$&  9.12$_{-0.86}^{+0.86}$&1.50 ( 20) \\           
s02                     &0.889                  & 0.40$_{-0.22}^{+0.20}$& 11.86$_{-5.18}^{+5.33}$& 4.59$_{-1.19}^{+1.18}$& 11.48$_{-1.02}^{+1.02}$& 4.76$_{-0.42}^{+0.42}$& 16.23$_{-1.45}^{+1.45}$&1.11 ( 25) \\           
s03,s04,s05,            &0.924                  & 0.40$_{-0.18}^{+0.17}$& 13.19$_{-4.97}^{+5.20}$& 5.89$_{-1.21}^{+1.19}$& 13.14$_{-0.96}^{+0.96}$& 5.95$_{-0.44}^{+0.44}$& 19.10$_{-1.40}^{+1.40}$&1.05 ( 93) \\           
s06,s30                 &                       &                       &                        &                       &                        &                       &                        &           \\     
ss07,s08,s09            &0.952                  & 0.54$_{-0.05}^{+0.05}$& 14.57$_{-1.38}^{+1.41}$& 6.69$_{-0.30}^{+0.30}$& 10.82$_{-0.20}^{+0.20}$& 6.60$_{-0.12}^{+0.12}$& 17.42$_{-0.33}^{+0.32}$&1.38 (423) \\           
s10                     &                       &                       &                        &                       &                        &                       &                        &           \\     
s11,s12,s13             &0.975                  & 0.56$_{-0.15}^{+0.14}$& 11.25$_{-3.76}^{+3.85}$& 7.74$_{-0.75}^{+0.74}$& 9.04$_{-0.40}^{+0.40}$ & 7.22$_{-0.32}^{+0.32}$& 16.26$_{-0.72}^{+0.72}$&1.14 ( 92) \\
s14, s31                &0.995                  & 0.77$_{-0.18}^{+0.17}$& 10.73$_{-3.67}^{+3.92}$& 7.11$_{-0.66}^{+0.65}$& 5.87$_{-0.26}^{+0.26}$ & 6.47$_{-0.29}^{+0.29}$& 12.34$_{-0.56}^{+0.55}$&1.15 ( 88) \\          
\hline
\end{tabular}
~\\
Notes: \emc ~and \emh ~are in units of $10^{56}$ cm$^{-3}$, \nhl ~is in units of 
$10^{22}$ cm$^{-2}$, and
\lxs, \lxh  ~and \lxt ~ are unabsorbed X-ray luminosities in soft, hard and 
total bands in units of 10$^{33}$ \lum. Abundances are in units of solar 
photospheric values.
$\chi^2_\nu$ is $\chi^2$ per degree of freedom and dof is the degree of freedom.
\end{center}
\end{table*}

\begin{figure}
\includegraphics[width=8cm]{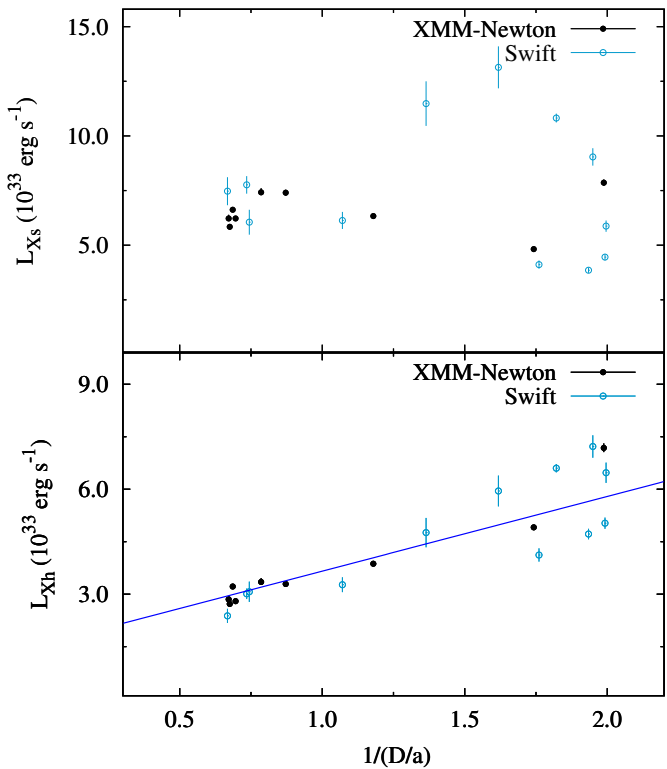}
\caption{X-ray luminosity as a function of inverse of binary separation in soft 
and hard X-ray bands. }
\label{fig:bsep}
\end{figure}

The EMs corresponding to the cool and hot temperatures change substantially, 
reflecting the variations in X-ray luminosities in the soft and hard bands. The 
X-ray luminosities observed for WR 25 were found to be  more than that for other 
close WR+O binaries and single WN stars \citep{zhekov12,skinner10}.   X-ray 
light curves of WR 25 as observed from \xmm ~ and \swift ~ show similar 
behavior. However, in terms of phase coverage, the light curves from \swift ~are 
much better than those observed from \xmm.  The deficit in X-ray flux just after 
periastron passage could be due to the eclipse of the wind interaction zone 
by the wind of the WN star.  In  all bands, the excess emission is  strongest 
near periastron passage. The X-ray  enhancement  after the eclipse  in the 
soft band indicates that besides the  colliding wind, individual components of the 
WR 25 system also contribute  to the X-ray emission. The soft X-ray flux 
peaked near  phase 0.5, which further supports that  both of the 
components of WR 25 are sources of soft X-rays, enhancing the combined soft X-ray 
flux when both of the stars are completely visible to the observer. During  
phase $\sim 0.7$, the soft X-rays further decrease to the minimum value 
indicating the possibility of a secondary eclipse, when the O-star is in front 
of the primary WN star. The deeper primary eclipse at phase $\sim 0.1$   could be due to the larger opacity of the WN wind.  On the other hand,  \lxh  ~was found to be at minimum during the phase 
$\sim 0.5$. Both  components of WR 25 are farthest apart at phase 0.5; therefore,    
the collision is weak, generating fewer X-rays.  This indicates that the hard 
X-rays originate from the  collision zone of the wind giving enhanced flux 
during periastron passage.  \cite{stevens92} and many other authors showed that 
strong winds from massive stars collide and generate hard X-rays, in addition to 
softer X-ray components due to intrinsic or individual components 
\citep{berghoefer97,sana06}.
The phase-locked X-ray variability could be a result of changing  separation in 
an eccentric orbit of WR 25. Fig \ref{fig:bsep} shows that the X-ray luminosity varies as a 
function of the inverse  binary separation [1/(D/a)]. It appears that soft X-ray 
luminosity does not depend on the binary separation. However, it is clearly seen 
that the hard X-ray luminosity is linearly dependent on the inverse of the binary 
separation. The Pearson correlation coefficients  for  \lxs ~versus 1/(D/a) and \lxh 
~versus 1/(D/a) were derived to be  -0.02 and 0.87 with a probability of no 
correlation found to be  0.94 and $10^{-5}$, respectively. This indicates that the 
hard X-rays in WR 25 are due to  the collision of winds. The wide massive binary 
systems whose X-ray luminosities follow the 1/D relation are  Cyg OB2 \# 9 
\citep{naze12} and HD 93205 \citep{antokhin03}.  Cyg OB2 \# 8A and WR140 
are the massive binary systems which  deviate from expectations at periastron, 
since the collisions then become radiative 
\citep{debecker06,pollock12,corcoran11}.  
However, there are a few wide binary 
systems that do not follow the 1/D variation e.g., WR 11 \citep{rauw00}, WR 22 
\citep{gosset09}. The present X-ray light curves of WR 25 strongly support the 
idea that during the rise in X-ray emission around  periastron passage the 
X-ray emission is primarily due to colliding winds \citep{willis95,stevens99}.

 Using V  $\sim8.03$ mag, a distance of $\sim 3.24$ kpc, a bolometric correction of 
-4.5 \citep{hamann06} a,nd an anomalous reddening of 4.5 mag \citep{hamann06}, the 
bolometric luminosity of WR 25 is calculated to be $1.25\times10^{40}$  \lum.  
The maximum and minimum values of \lxb ~for WR 25 are thus calculated to be 
$10^{-5.81}$ and $10^{-6.18}$, respectively. The derived value of \lxb ~during 
periastron passage is similar to that derived by \cite{seward82} using 
Einstein observations. During the  phases $\sim 0.03$ and $\sim 0.7$, the 
possible primary and secondary eclipse positions where the WN and O-type stars 
contributed more in soft X-rays, respectively, the \lxb  ~are calculated to be 
$10^{-6.5}$ and $10^{-6.31}$, respectively.  These values of \lxb ~are similar 
to those obtained for other single WR and O-type stars \citep{skinner10,sana06}.  

 The hard X-rays from WR 25 appear to be a result of  collision of winds from binary 
companions. The gas in the colliding wind regions could  either be adiabatic or 
radiative depending on the  cooling parameter ($\chi_c$) as $\chi_c = 
v_\infty^4 d/ \dot{M}$ \citep{stevens92}, where $v_\infty$ is in units of  10$^3$ 
km s$^{-1}$, $d$ is the distance from the contact to star in  units of  $10^7$ 
km and $\dot{M}$ is in units of $10^{-7}$ \msun yr$^{-1}$. For $\chi_c \gtrsim 
1$, the wind can be assumed to be adiabatic while for  $\chi_c \ll 1$, it is 
roughly isothermal. The cooling parameter is directly proportional to the 
distance of contact ($d$); as a consequence of this,  $\chi_c \propto 
P_{orb}^{2/3}$ \citep{stevens92}. This means that the shocked region will be 
adiabatic for  longer period binaries.  WR 25 is  one of the longer orbital 
period binaries, indicating the shocked region is adiabatic. Furthermore, if we 
assume that the contact of winds is close to the O-type star of WR 25, the value of 
$\chi_c$ is calculated to be $> 1$, indicating an adiabatic wind in the shocked 
region.  In our analysis, the abundances of O, Ne, Mg, Si, S, Ca, Ar, Fe and Ni 
were found to be non-solar, which is expected from WR stars because they are in 
advanced evolutionary stages.  The non-solar abundances of these elements could 
be due to the presence of hot plasma near the surface of the O4 star,  due 
to the collision of strong supersonic winds from the WN6 star with the less powerful 
supersonic wind from O4 star.   For wide binaries, \cite{pittard02} showed 
that WR winds dominate the X-ray luminosity.  \cite{luo90} and 
\cite{myasnikov93} have also found  that the shocked WR stellar wind dominates 
the X-ray emission for  WR+O systems in the adiabatic limit.
 Furthermore, the shocked volume and the emission measure near the O-type star 
may be  dominated by WN winds; therefore, the X-ray emitting plasma also shows 
non-solar  abundances. When we compared the derived values of abundances of 
different elements for WR 25 with other WN-type stars, we found that the 
abundances of Ne and S in WR 25 are similar to that of WN-type stars 
\citep{smith05,ignace07}

The light curve of WR 25 in the V band shows minimum 
light near phase 0.75, which could be possible position of eclipse  when the WN star 
is behind the O-star.  It appears that the WN star is brighter than the O-type star 
in the WR 25 system; therefore, may  observe an eclipse near phase 0.75.   
\cite{hamann06} also showed that O-type stars contribute a  minimum of  15\%    
to the total light of the system in the V band and  \cite{gosset94} had also noticed 
the photometric variation in WR 25 with a variability amplitude of 0.02 mag in 
the Str\"{o}mgren b-band on a time scale of $\sim$250 days. Many other WR binaries 
also show  phase-locked variability in the optical band and the variability is 
attributed to their binary  nature \citep{lamontagne96}. The presence of eclipses 
in X-ray and optical light curves indicates that the orbital plane of  WR 25 
has a high inclination angle.  The minimum of the V-band light curve is consistent with the phase minimum at $\sim 0.7$ of the soft X-ray light curve. However, unlike to the soft X-ray light curve we could not see any other minimum in the V-band light curve near phase 0.03, where \nhl ~was also found to be maximum. This could be  due to the poor phase coverage of the V-band light curve.

The analysis of the  present data shows that WR 25 is a colliding wind binary with 
an orbital period of $\sim 208 $ days, where the hard X-rays could be due to the 
collision zone while soft X-rays could be attributed to individual components.

\acknowledgments
We thank the referee for his/her  comments and suggestions that helped to considerably improve 
the manuscript.  This work uses data obtained by XMM-Newton, an ESA 
science mission with instruments and contributions directly funded by ESA Member 
States and the USA (NASA). We acknowledge the Swift data archive and the UK Swift Science Data Center.

{\it Facilities:} \facility{\xmm}, \facility{\swift}.

\newpage

\end{document}